\documentclass[a4paper,11pt]{article}
\usepackage{jinstpub} 
\usepackage{fancyhdr} 
\usepackage{siunitx}
\usepackage{textcomp}
\usepackage{mathtools}
\usepackage{dblfloatfix} 
\usepackage{lineno}
\usepackage{todonotes}
\usepackage{float}

\title{\boldmath Design, Construction, and Testing of the APOLLO ATCA Blades for Use at the HL-LHC}
\author[a]{Alp Akpinar,}
\author[a]{Aymeric Blaizot,}
\author[a]{Serhii Cholak,}
\author[a]{Gianfranco de Castro,}
\author[a]{Zeynep Demiragli,}
\author[b]{Alec Duquette,}
\author[a,1]{Jonathan Richard Fulcher,\note{speaker and corresponding author}}
\author[a]{Dan Gastler,}
\author[c]{Kristian Hahn,} 
\author[a]{Eric Shearer Hazen,}
\author[a]{Si Hyun Jeon,}
\author[b]{Peace Kotamnives,}
\author[a]{Alexander Madorsky,}
\author[c]{David Monk,}
\author[c]{Sheena Noorudhin,}
\author[b]{Michael Oshiro,}
\author[a]{James Rohlf,}
\author[b]{Charles Ralph Strohman,}
\author[d]{Emily Minyun Tsai,} 
\author[b]{Peter Wittich,}
\author[a]{Siqi Yuan,}
\author[b]{and Rui Zou}

\emailAdd{jonathan.fulcher@cern.ch}

\affiliation[a]{Department of Physics, Boston University, Commonwealth Ave, Boston, MA, U.S.A.}
\affiliation[b]{Department of Physics, Cornell University, Sciences Dr, Ithaca, NY, U.S.A.}
\affiliation[c]{Department of Physics, Northwestern University, Sheridan Rd, Evanston, IL, U.S.A.}
\affiliation[d]{Department of Physics, Northeastern University, Forsyth St, Boston, MA, U.S.A.}

\abstract{The Apollo Advanced Telecommunications Computing Architecture (ATCA) platform is an open-source design consisting of a generic “Service Module” (SM) and a customizable “Command Module” (CM), allowing for cost-effective use in applications such as the readout of the inner tracker and the Level-1 track trigger for the CMS Phase-II upgrade at the HL-LHC. The SM integrates an intelligent IPMC, robust power entry and conditioning systems, a powerful system-on-module computer, and flexible clock and communication infrastructure. The CM is designed around two Xilinx Ultrascale+ FPGAs and high-density, high-bandwidth optical transceivers capable of 25 Gb/s. Crates of Apollo blades are currently being tested at Boston University, Cornell University, and CERN.}


\arxivnumber{2501.03702} 

\collaboration[c]{on behalf of the CMS Collaboration}

\proceeding{Topical Workshop on Electronics for Particle Physics TWEPP2024 \\
  September 30 - October 4, 2024\\
  Online}

\notoc

\begin{document}
\maketitle
\newpage 
\pagenumbering{arabic}
\flushbottom
\hrule
\tableofcontents
\vspace{0.5cm} 

\hrule
\section{Introduction}
\label{sec:intro}

High-performance Advanced Telecommunications Computing Architecture (ATCA) blades~\cite{Apollo2019} are becoming essential in high-energy physics experiments, particularly for the High-Luminosity LHC. The Apollo platform, aptly named after the Apollo space program\cite{APOLLO}, addresses challenges in power, cooling, optical fiber management, and communication interfaces by providing a streamlined hardware environment along with firmware and software tool kits. It features a versatile “Service Module” (SM) that manages ATCA communications, power distribution, and clocking, paired with an application-specific “Command Module” (CM) equipped with FPGAs and numerous optical links for high data rates (see section \ref{sec:REV2_TESTS}).

The SM is a standard-sized ATCA blade that supports a CM within a 7U $\times$ 180 mm cutout. It includes an Enclustra Zynq Ultrascale+ System-on-Module (SoM) with an embedded Linux OS for control and monitoring, along with an Intelligent Platform Management Controller (IPMC), OpenIPMC~\cite{OpenIPMC}, and a Wisconsin Ethernet Switch Module~\cite{ESM}. The SM provides flexible interfaces such as gigabit Ethernet, additional Ethernet to the switch, and four 10 Gb/s bidirectional serial links to the CM, along with diagnostic and programming interfaces like asynchronous serial, I\textsuperscript{2}C, and JTAG. Power delivery is managed by commercial modules, supplying up to 12 V$_\text{DC}$ at 30 A, while the platform supports more than 100 optical links at speeds up to 25 Gb/s to meet the high data throughput demands of experiments. 

\section{Overview and CMS Applications}
\label{sec:overview}

The CMS experiment will undergo major upgrades for the High-Luminosity LHC~\cite{CMSDetector}, including a new silicon tracker~\cite{CMSTracker}. The Apollo platform will be used in four CMS subsystems: the Level-1 Track Finder (TF), the Inner Tracker Data Trigger and Control system (IT-DTC), the Precision Proton Spectrometer (PPS), and the high-precision luminosity monitoring system (BRIL)~\cite{BRIL}.

The Level-1 TF reconstructs tracks from the Outer Tracker and calculates track parameters for the Level-1 trigger system. Although the algorithm requires substantial FPGA resources, the total power consumption should remain within operating limits (< 100 W). It requires two Virtex Ultrascale+ FPGAs and 62 optical links at 25\,Gb/s and a total of 162 Apollo blades. 

\begin{figure}[H] 
    \centering
    \includegraphics[width=1\linewidth]{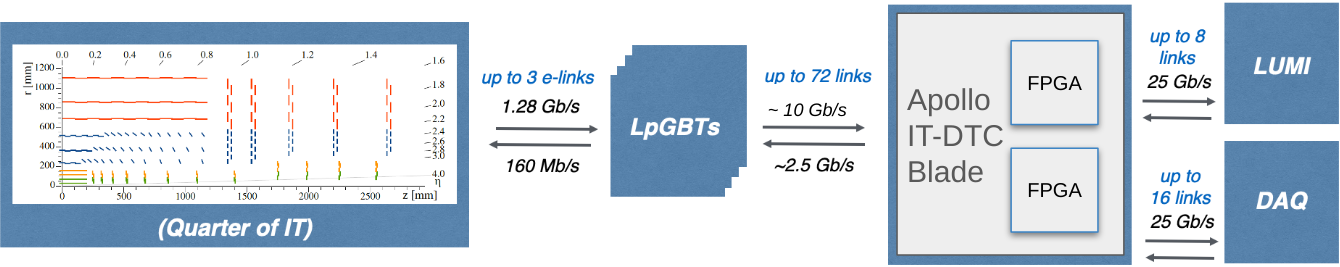}
    \vspace{-25pt} 
    \caption{Example CMS Application IT-DTC.}
    \label{fig:IT-DTC}
\end{figure}

The IT-DTC reads hits from the front-end pixel ASIC chips and converts the data into a compact format. It also converts the trigger signal from the CMS trigger to tokens and transmits them to the front-end. It requires up to 72 LpGBT~\cite{LPGBT} optical links at 10\,Gb/s, 16 optical links at 25\,Gb/s for central data acquisition (CDAQ), 4 optical links on each of the TEPX DTCs at 25\,Gb/s for BRIL "LUMI" and totals 36 Apollo boards (figure~\ref{fig:IT-DTC}).

The BRIL Lumi system receives event fragments on 4 to 8 25 Gb/s transceivers per blade from IT-DTC with dedicated luminosity processing hardware. These 25 “luminosity blade” Apollos perform real-time pixel clustering and cluster counting on CM FPGAs. The data are then decoded and processed. 
Additionally, the PPS also made a recent request for three Apollos.

To satisfy the requirements of these applications, the CMS version of the Apollo CM can accommodate one or two large Xilinx Ultrascale+ FPGAs with 4-channel transceiver and 12-channel transmitters and receiver Firefly optical engines that total 112 optical links with speeds up to 25\,Gb/s. Two revisions have been tested, and a third revision is under development, of which the first three blades should be assembled in Q4 of 2024. Revision~2 (Rev2) was produced in Q3 of 2021 and has undergone thorough testing and evaluation. This paper describes the evaluation, shares what we learned, presents the performance of the Rev2 design (section \ref{sec:REV2_TESTS}), and discusses the major updates and the test status of the revision~3 (Rev3) design (section \ref{sec:REV3_TESTS}).

\section{Revision 2 SM and CM Performance Tests and Results}
\label{sec:REV2_TESTS}
\begin{figure*}
    \vspace{-10pt} 
    \begin{minipage}{0.53\textwidth}
        \centering
        \vspace{-10pt} 
        \includegraphics[width=0.88\linewidth,height=0.67\linewidth]{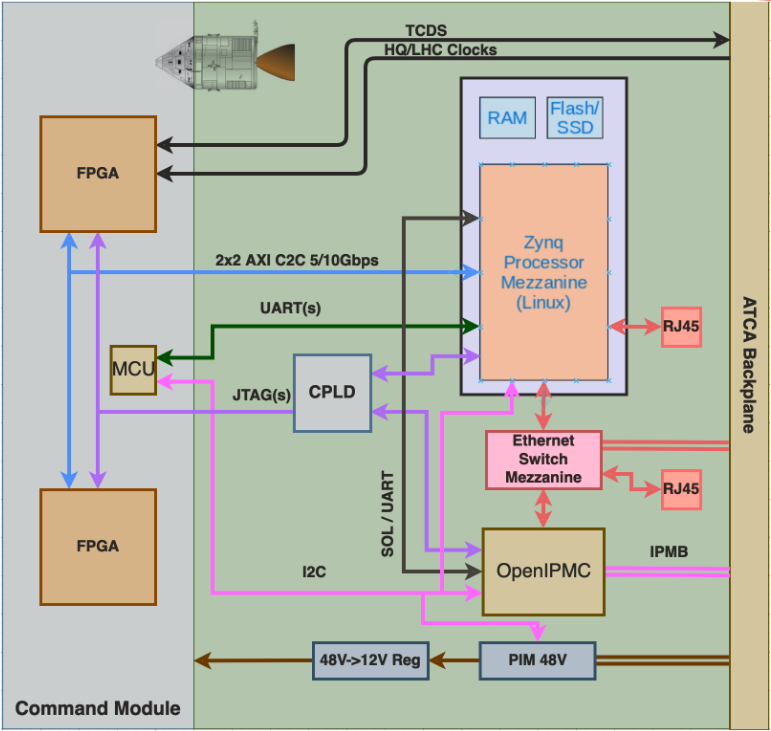}
        \vspace{-10pt} 
        \caption{Apollo schematic.}
        \label{fig:apollo} 
    \end{minipage}
    \qquad
    \begin{minipage}{0.47\textwidth}
        \centering
        \vspace{-10pt} 
        \includegraphics[width=0.95\textwidth]{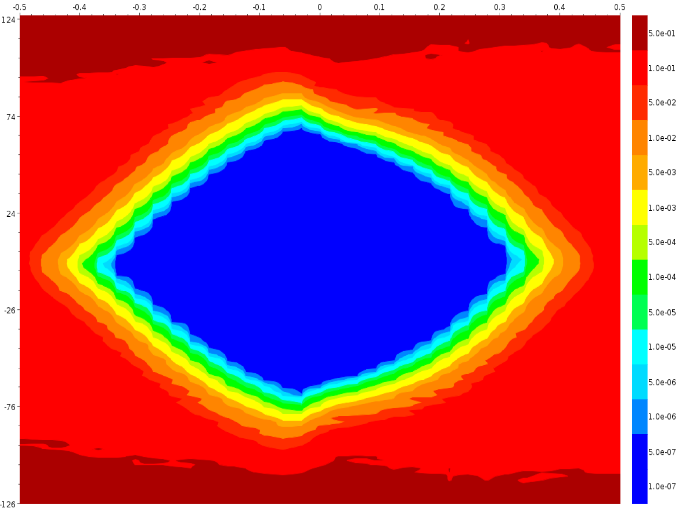}
        \vspace{-10pt} 
        \caption{Typical eye with large opening at 10\,Gb/s over TCDS lines from backplane on the Rev3 SM.}
        \label{fig:sm_tcds_eye}
    \end{minipage}
\end{figure*}

The CMS Rev2 Apollo blade (figure~\ref{fig:apollo}) has two large FPGAs (XCVU13P, A2577 package) and sites for four- and twelve-channel Firefly optical transceivers with up to 124 links supporting speeds up to 25\,Gb/s. A comprehensive set of performance tests was conducted on Rev2 to validate data transfer integrity, clock distribution, and microcontroller (MCU) code performance. Thorough thermal tests first presented in~\cite{Apollo2021} were also repeated. This section summarizes the testing procedure and the final results.

\subsection{Hardware and Firmware Tests}
\label{sec:HW_&_FW_TESTS}
Link integrity tests focused on high-speed data transfer across FireFly optical links and FPGA interconnects, achieving bit error rates (BER) less than $10^{-16}$ at 25 Gb/s and 10 Gb/s. Extensive testing of the CMS Standard Protocol (CSP)\cite{EMP} links was performed. Copper links were tested in single-FPGA loop-back and cross-FPGA configurations at both Boston University (BU) and CERN.  Similarly, the 4-channel-transceiver FireFly optical modules were tested in single-FPGA loop-back and cross-FPGA configurations, along with 12-channel 25Gb/s FireFly modules with single-FPGA-loop-back and reverse-loop-back tests at CERN. All tests used both pseudo-random binary sequence (PRBS), and CSP with the CMS Clock and Trigger Distribution System (TCDS2)\cite{TCDS2} modes. The CSP tests, which required correct timing alignment, were monitored and controlled through the CLI and Apollo Herd interfaces (see section~\ref{sec:SOFTWARE_FRAMEWORK_TESTS}). Chip-to-chip (C2C) links between the Zynq and FPGAs were also validated at 5Gb/s. Advanced eXtensible Interface (AXI) C2C link tests were conducted from the Zynq Linux OS to endpoints in the SM and CM, with the goal of evaluating data integrity and write speed. Although the current AXI C2C links do not employ Direct Memory Access (DMA), the test utility supported a variety of block size configurations, and dedicated firmware with Block-RAM (BRAM) was developed to support block write tests. Extended tests showed no errors over 96 hours (343 GB) and 14 hours (47 GB), respectively. However, the maximum write speed achieved was 24 Mb/s with block sizes larger than 4096 words. Future plans include developing an AXI DMA driver to increase transfer speeds closer to the selected physical link rate.
The clock distribution test was performed using TCDS2, integrated within the CMS back-end system firmware framework (EMP)\cite{EMP}. The test involved receiving clock and data stream signals from a DTH\cite{DTH} via the backplane to FPGA1 on a CM, with a secondary relay to the second FPGA. Version 0.1.1 of the TCDS2 firmware successfully verified the lightweight endpoint on both FPGAs and demonstrated proper recovery of the 40 MHz LHC clock signal from the TCDS2 stream along with successful operation of the TCDS2 relay. More advanced testing with the latest version 0.2.0rc1 introduced the full endpoint and confirmed recovery of the 40 MHz clock stream on both FPGAs, with successful testing of both the internal and redistributed LHC 40 MHz and 320 MHz reference clocks. Further development is ongoing for an upgraded inter-FPGA relay compatible with the latest TCDS2 firmware.

\begin{figure}
    \centering
    \vspace{-15pt} 
    \includegraphics[width=1\linewidth]{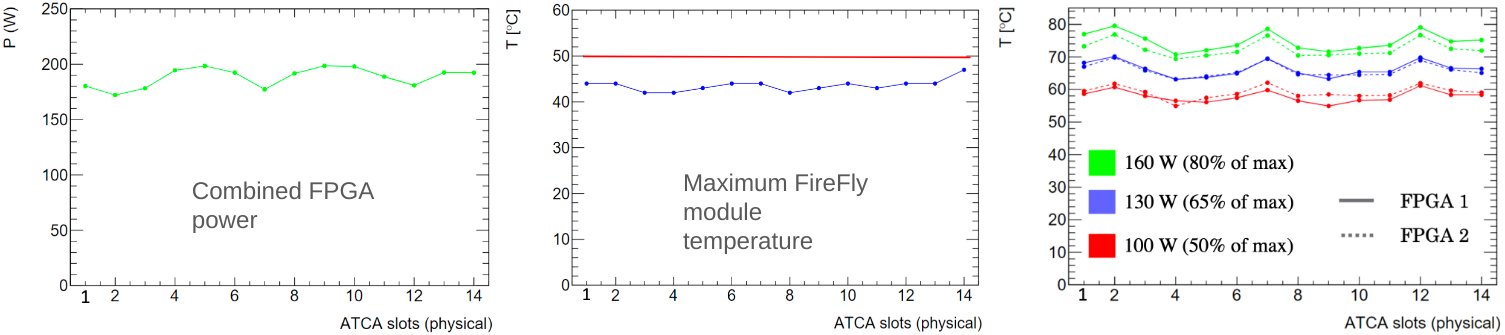}
    \vspace{-30pt} 
    \caption{\label{fig:finalperf}Temperature vs power as a function of ATCA slot for both VU13P FPGAs, and FireFly module compared to the recommended operating range of below 50~$^\circ$C, of the Rev2 CM.}
    \label{fig:thermal}
\end{figure}

Thermal performance tests were conducted at CERN to evaluate power and cooling efficiency (see figure~\ref{fig:thermal}). The test setup involved fixing the fan speed to 10/15 (67\% of maximum), maintaining an inlet air temperature of 23°C, and running the heater firmware\cite{heaterpaper} on both FPGAs. The heaters were gradually activated until both FPGAs reached 85°C. At this point, the power consumption of the FPGAs and the maximum temperature of the FireFly modules were recorded, with the latter staying below 50°C (47°C maximum), demonstrating minimal thermal coupling to the FPGAs. Furthermore, FPGA temperatures were measured at three more realistic combined power levels, and the tests were repeated in different slots to verify uniform behavior throughout. The airflow impedance was found to be slightly high (0.26 in-H$_2$O @ 30 CFM). However, the system was able to accommodate up to 200W of FPGA power with a split of approximately 55\% to FPGA1 and 45\% to FPGA2, confirming the platform’s ability to manage significant thermal loads effectively (see figure~\ref{fig:thermal}). The MCU code was also tested, ensuring the proper configuration and sequencing of power supplies, as well as the correct setup and monitoring of FireFly modules and clocking circuits. Furthermore, environmental data was successfully transmitted to both the IPMC and the Zynq, ensuring robust control and monitoring of the system.

Lastly, the platform passed a significant milestone by validating the LpGBT slow control and data acquisition front-end link, which will be used in the Inner Tracker upgrade, by exercising the link to write and read registers on a front-end module. This test demonstrated that the Apollo platform is capable of handling the communications necessary for future CMS applications.




\subsection{Software Framework Tests}
\label{sec:SOFTWARE_FRAMEWORK_TESTS}

The Apollo platform integrates seamlessly with the CMS Online Software Framework, known as EMP Shep/Herd, via a plugin called Apollo Herd, which provides a set of explicit commands to control and monitor various aspects of the system. These commands include powering on/off and programming the FPGAs, and running algorithm or link integrity tests. The software also allows direct access to Apollo's address tables through a custom command line interface (CLI) called “BUTool”, providing read/write register access and facilitating low-level control. Several abstracted objects are utilized within the framework, including the SM and MCU, for monitoring status tables from BUTool; the CM FPGAs, for programming and conducting link tests; and both copper links and FireFly optical modules, serving as link test endpoints and monitoring targets. 

Monitoring functions include real-time status tables, FireFly module temperatures, and optical power levels, as well as key FPGA parameters such as firmware versions, build timestamps, clock frequencies, and temperatures; Ensuring a robust environment for both control and monitoring.

\section{Revision 3 SM Performance Tests and CM Plans}
\label{sec:REV3_TESTS}
The Rev3 SM underwent a series of tests after its initial receipt in July of this year. Changes in Rev3 relative to Rev2 are minimal and consist of streamlining the routing of the TCDS2 link to improve link quality by eliminating unnecessary switching. Initial testing focused on verifying the basic functionality of three units, starting with resistance and power measurements to ensure proper electrical behavior. The Ethernet switch and IPMC were evaluated to confirm stability. Ethernet connectivity was also tested to ensure seamless network integration. Furthermore, the Xilinx Zynq SoC was tested, including successful boot of the operating system (OS). A link eyescan was performed to validate TCDS2 signal integrity, and the new over-voltage and under-voltage protection circuits were thoroughly tested to ensure they functioned correctly within the specified limits. The over-voltage condition was simulated by changing the ratio of the resistive divider in the sensor circuit. The under-voltage condition was tested by simply turning the payload DC-DC converter off and checking that the sensor had tripped at the appropriate voltage using a scope. The sensors readout from the SM was also tested to verify that all monitored parameters were accurately reported.
In parallel, the Rev3 SM was tested in combination with the Rev2 CM. These tests included programming of both FPGAs via JTAG, as well as read/write operations over the C2C AXI interface, and similar read/write operations over IPBUS, to both FPGAs. C2C link tests were conducted to assess the integrity and performance of the connections between the SM and CM. The connection to the MCU was also tested, along with sensor readouts from the CM to ensure comprehensive monitoring capabilities. These tests demonstrated the 
compatibility and communication between the revised SM and the older CM.
%

Several steps were completed to ensure the SM’s power delivery system functions correctly. The testing process involved connecting the IPMC module and enabling payload power through a jumper. The Complex Programmable Logic Device (CPLD) was then programmed, followed by a power-up check to verify that the IPMC had correctly enabled payload power. After this, the SoC was connected, and power was checked to confirm that all required voltages were properly enabled. Additionally, the Rev2 CM was connected to facilitate further testing.

After installing the SM in the crate, Ethernet access was verified and the TCDS2 system was checked.  Integrated Bit Error Ratio Test (IBERT) tests were conducted on the TCDS2 backplane link to assess the performance between the DTC and the Apollo SM, with the quality of the signal plotted against the crate slot. The worst-case scenario was evaluated with a long-term test in which the BER was measured as far as $2 \times 10^{-15}$, with no error detected. Eye-plot shapes were checked and no error was seen. The results confirmed reliable communication between the Rev3 SM and the DTH using PRBS-31 patterns, aided by Decision Feedback Equalization (DFE); see figure~\ref{fig:sm_tcds_eye}.

\section{Revision 3 Preproduction and Plans}
\label{sec:REV3_PLANS}

The strategy for Apollo preproduction is to synchronize across all systems to reduce procurement and production costs. A pilot production run of approximately 30 boards is anticipated, primarily for the TF, with expected completion in Q1 2025. Production quantities include 162 TF boards plus 10 spares, 36 IT-DTC boards plus 5 spares, and 16 BRIL boards plus 4 spares, with the bulk expected in Q3 of 2026. Following completion of the currently ongoing Rev3 SM testing, full pre-production numbers will be initiated. The plan includes production of all PCB boards for the CM. Three boards will be populated, one without FPGAs. These boards will be tested with the final production versions of the Samtec Firefly 12-channel, 25 Gb/s optical parts once they become available in early 2025. Thorough testing of the Rev3 SM and CM is needed due to layout changes, most notably the addition of two inter-FPGA copper links, bringing the total to 56, and an improvement to the distribution and layout of data and clock signals for TCDS2, adding more choice options for the master LHC clock source, while enhancing the symmetry between the two FPGAs, thus minimizing firmware design differences. For Rev3 SM changes, see section \ref{sec:REV3_TESTS}.

\section{Summary}
The Apollo ATCA blade will be used for the IT-DTC, TF, PPS, and BRIL systems of the CMS experiment at the High-Luminosity LHC (HL-LHC). Rev3 of the designs for the SM and CM are now finalized for preproduction. The first two revisions have demonstrated excellent link integrity, thermal performance, and adaptability to various requirements. Currently, Apollo Rev2 is being used for development and testing at multiple locations, including Boston University (BU), Cornell, and CERN. The SM Rev3 has been produced and is being tested at BU, while the CM Rev3 is nearing manufacture. The validation of the SM board is complete for Rev2 and is ongoing for Rev3.

The SM Zynq firmware has undergone extensive testing, while testing of the CM TCDS2 clocking scheme is well advanced, leading to high confidence in the final design. The CMS application framework, EMP, is also  maturing, with testing being conducted for IT-DTC, TF and BRIL. Initial testing of the IT-DTC DAQ chain is scheduled at CERN in Q4 of 2024.
Software CLI tools are very mature and well tested, with the EMP board-level plugin nearly complete and currently under test. Application-specific EMP plugins will follow.


\acknowledgments
We would like to thank the authors of ref.~\cite{heaterpaper} for the 
code 
used in our power tests. This work was supported by the National Science Foundation under grants NSF-PHY-1946735, 
NSF-PHY-2209443, 
and NSF-PHY-2046626 
and by the Department of Energy under grants 
DE-SC0021174 and 
DE-SC0016021. 






\end{document}